\documentclass[aps,prd,showpacs,superscriptaddress]{revtex4}
\usepackage{graphicx}
\usepackage{bm}
\begin{document}
\title{Third-order perturbative solutions in the Lagrangian
perturbation theory with pressure}
\author{Takayuki Tatekawa}
\email{tatekawa@gravity.phys.waseda.ac.jp}
\affiliation{Department of Physics, Waseda University,
3-4-1 Okubo, Shinjuku-ku, Tokyo 169-8555, Japan}
\date{\today}

\begin{abstract}
Lagrangian perturbation theory for cosmological fluid describes
structure formation in the quasi-nonlinear stage well. We present
a third-order perturbative equation for Lagrangian perturbation
with pressure in both the longitudinal and transverse modes. Then
we derive the perturbative solution for simplest case.
\end{abstract}


\maketitle

\section{Introduction}\label{sec:intro}

The structure formation scenario based on gravitational instability
has been studied for a long time. The Lagrangian perturbative
method for the cosmological fluid describes the nonlinear
evolution of density fluctuation rather well. 
Zel'dovich~\cite{zel} proposed
a linear Lagrangian approximation for dust fluid.
This approximation is called the Zel'dovich approximation
(ZA)~\cite{zel,Arnold82,Shandarin89,buchert89,coles,saco}.
ZA describes the evolution of density fluctuation better than
the Eulerian approximation~\cite{Munshi94,Sahni96,yoshisato}.
After that, the second- and the third-order perturbative
solution for dust fluid were derived
\cite{Barrow93,Bouchet92,Bouchet95,Buchert92,Buchert93,Buchert94,Catelan95,Sasaki98}.

Recently the effect of the pressure in the cosmological
fluid has been considered. At first, the effect of the pressure
is originated from velocity dispersion using the
collisionless Boltzmann equation~\cite{BT, budo}.
Buchert and Dom\'{\i}nguez~\cite{budo} showed that
when the velocity dispersion is regarded
as small and isotropic it produces effective "pressure" or
viscosity terms. Furthermore, they posited the
relation between mass density $\rho$ and pressure $P$, i.e.,
an "equation of state". 
Adler and Buchert~\cite{Adler99} have formulated the
Lagrangian perturbation theory for a barotropic fluid.
Morita and Tatekawa~\cite{Morita01} and Tatekawa
\textit{et al.}~\cite{Tatekawa02}
solved the Lagrangian perturbation equations for a polytropic fluid
up to the second order.
Hereafter, we call this model the "pressure model".

Although the higher-order perturbative solution is expected
to improve the approximation, there is a counterargument.
Let us consider the evolution of the spherical void for dust fluid.
Because the exact solution has already been derived, we can discuss
the accuracy of the perturbative solutions. In this case, when we increase
the order of Lagrangian approximation, contrary to expectation,
the description becomes worse~\cite{Munshi94,Sahni96,Tatekawa05}.
Especially, when we stop the order of the perturbation
until even order (the second-order), 
the perturbative solution describes the contraction of a void
at a later.

For the pressure model, although we do not know the exact solution
for the evolution of the spherical void, the same problem may arise.
In fact, according to a comparison between the first-order
solution and the full-order numerical solution for the spherical
symmetric model, the difference
of these solutions obviously appears at a late time
(density fluctuation $\delta >1$ or $\delta <-0.5$)
~\cite{Tatekawa04B}.

In this paper, we present the third-order perturbative equation
for the pressure model. Then we derive the perturbative solution
for the simplest case, i.e., where the background is given by 
the Einstein-de Sitter (E-dS) Universe model and the polytropic index
$\gamma=4/3$. Furthermore, we compare the evolution of the
density fluctuation between the first-, second-, and third-order
approximations for one-dimensional model.

This paper is organized as follows.
In Sec.~\ref{sec:basic}, we present a Lagrangian description of
basic equations for cosmological fluids.
In Sec.~\ref{sec:Perturbation}, we show perturbative equations
and derive perturbative solutions for the pressure model
up to the third order.
In Sec.~\ref{subsec:1st-eq}, we show the perturbative equations
and solutions of the first- and the second-order perturbation.
For the higher-order perturbation, we ignore the first-order
transverse modes.
In Sec.~\ref{subsec:3rd-eq}, we present the perturbative equations
for the third-order approximation. In general, it is extremely hard
to solve the third-order perturbative equations. Therefore in
Sec.~\ref{subsec:3rd-sol}, we derive the solutions for the simplest
case (E-dS Universe model, $\gamma=4/3$).

In Sec.~\ref{sec:1D}, we introduce spacial symmetry. We
consider a planar model and compare the evolution of the
density fluctuation between the first-, second-, and third-order
approximations.
In Sec.~\ref{sec:summary}, we summarize our conclusions.

In Appendix~\ref{sec:2nd-sphe}, we present the second-order
perturbative equation for a spherical symmetric model in
the pressure model. Because of mode-coupling of the first-order
perturbations, it seems difficult to solve.

\section{Basic equations}\label{sec:basic}

Here we briefly introduce the Lagrangian description for cosmological
fluid. In the comoving coordinates,
the basic equations for cosmological fluid are described as
\begin{eqnarray}
\frac{\partial \delta}{\partial t} + \frac{1}{a} \nabla_x \cdot \{ \bm{v}
(1+\delta) \} &=& 0 \,, \label{eqn:comoving-conti-eq} \\
\frac{\partial \bm{v}}{\partial t} + \frac{1}{a} (\bm{v} \cdot \nabla_x)
\bm{v} + \frac{\dot{a}}{a} \bm{v} &=& \frac{1}{a} \tilde{\bm{g}} - \frac
{1}{a \rho} \nabla_x P \,, \label{eqn:comoving-Euler-eq} \\
\nabla_x \times \tilde{\bm{g}} &=& \bm{0} \,, \label{eqn:rot-g} \\
\nabla_x \cdot \tilde{\bm{g}} &=& - 4 \pi G \rho_b a \delta \,,
\label{eqn:comoving-Poisson-eq} \\
\delta & \equiv & \frac{\rho - \rho_b}{\rho_b} \,.
\end{eqnarray}
In the Eulerian perturbation theory, the density fluctuation $\delta$
is regarded as a perturbative quantity. On the other hand, in the
Lagrangian perturbation theory, the displacement from homogeneous
distribution is considered.
\begin{equation} \label{eqn:x=q+s}
\bm{x} = \bm{q} + \bm{s} (\bm{q},t) \,,
\end{equation}
where $\bm{x}$ and $\bm{q}$ are the comoving Eulerian coordinates
and the Lagrangian coordinates, respectively. $\bm{s}$ is
the displacement vector that is regarded as a perturbative quantity.
From Eq.~(\ref{eqn:x=q+s}), we can solve the continuous equation
Eq.~(\ref{eqn:comoving-conti-eq}) exactly. Then the density
fluctuation is given in the formally exact form.
\begin{equation}
\delta = 1 - J^{-1}, ~~ J \equiv \det \left (
 \frac{\partial x_i}{\partial q_j} \right ) \,.
\end{equation}
$J$ means the Jacobian of the coordinate transformation
from Eulerian $\bm{x}$ to Lagrangian $\bm{q}$.
Therefore, when we derive the solutions for $\bm{s}$, we can know
the evolution of the density fluctuation. 

The peculiar velocity is given by
\begin{equation}
\bm{v}=a \dot{\bm{s}} \label{eqn:L-velocity} \,.
\end{equation}
Then we introduce the Lagrangian time derivative:
\begin{equation}
\frac{\rm d}{{\rm d} t} \equiv \frac{\partial}{\partial t}
 + \frac{1}{a} \bm{v} \cdot \nabla_x \,. \label{eqn:dt-L}
\end{equation}
Taking divergence and
rotation of Eq.~(\ref{eqn:comoving-Euler-eq}), we obtain
the evolution equations for the Lagrangian displacement:
\begin{eqnarray}
\nabla_x
 \cdot \left (\ddot{\bm{s}}
  + 2 \frac{\dot{a}}{a} \dot{\bm{s}}
  - \frac{\kappa \gamma \rho_b^{\gamma-1}}{a^2}
 J^{-\gamma} \nabla_x J \right )
&=& -4 \pi G \rho_b (J^{-1} -1) \,, \label{eqn:L-longi-eqn} \\
\nabla_x \times \left (\ddot{\bm{s}} + 2 \frac{\dot{a}}{a}
 \dot{\bm{s}} \right )
&=& 0 \,, \label{eqn:L-trans-eqn}
\end{eqnarray}
where $(\dot{})$ means the Lagrangian time derivative (Eq.~(\ref{eqn:dt-L})).
To solve the perturbative equations,
we decompose the Lagrangian perturbation to the longitudinal
and transverse modes:
\begin{eqnarray}
\bm{s} &=& \nabla S + \bm{s}^T \,, \\
\nabla \cdot \bm{s}^T &=& 0 \,,
\end{eqnarray}
where $\nabla$ means the Lagrangian spacial derivative.

Here we expand the Jacobian:
\begin{eqnarray}
J &=& 1 + s_{i,i} + \frac{1}{2} \left ( s_{i,i} s_{j,j}
 - s_{i,j} s_{j,i} \right ) + \det \left (s_{i,j} \right ) \\
  &=& 1 + \nabla^2 S + \frac{1}{2} \left \{ (\nabla^2 S)^2
  - S_{,ij} S_{,ji} - s^T_{i,j} s^T_{j,i} 
  - 2 S_{,ij} s^T_{j,i} \right \} \nonumber \\
  &&  
  + \det \left (S_{,ij} + s^T_{i,j} \right ) \,.
\end{eqnarray}
Because Eqs.~(\ref{eqn:L-longi-eqn}) and (\ref{eqn:L-trans-eqn})
include the Eulerian spacial derivative, we change to the Lagrangian
spacial derivative.
\begin{eqnarray*}
\frac{\partial}{\partial x_i} &=& \frac{\partial}{\partial q_i}
 - s_{j,i} \frac{\partial}{\partial x_j} \\
 &=& \frac{\partial}{\partial q_i}
 - s_{j,i} \frac{\partial}{\partial q_j}
 + s_{j,i} s_{k,j} \frac{\partial}{\partial x_k} \\
 &=& \frac{\partial}{\partial q_i}
 - s_{j,i} \frac{\partial}{\partial q_j}
 + s_{j,i} s_{k,j} \frac{\partial}{\partial q_k} + \cdots \,.
\end{eqnarray*}

\section{The Lagrangian perturbative solutions}
\label{sec:Perturbation}

\subsection{The first- and second-order perturbative solutions}
\label{subsec:1st-eq}

From Eqs.~(\ref{eqn:L-longi-eqn}) and (\ref{eqn:L-trans-eqn}),
we obtain the first-order perturbative equations:
\begin{eqnarray}
\nabla^2 \left ( \ddot{S}^{(1)} + 2 \frac{\dot{a}}{a}
 \dot{S}^{(1)} - 4\pi G \rho_b S^{(1)}
  - \frac{\kappa \gamma \rho_b^{\gamma-1}}{a^2} \nabla^2 S^{(1)}
 \right ) &=& 0 \,, \\
\nabla \times \left ( \ddot{\bm{s}}^{T (1)}
 + 2 \frac{\dot{a}}{a} \dot{\bm{s}}^{T (1)}
\right  ) &=& {\bm{0}} \,.
\end{eqnarray}
The first-order solutions for the longitudinal mode depend on
spacial scale. Therefore the solutions are described with a
Lagrangian wavenumber. In this paper, we discuss only perturbative
solutions in the E-dS Universe model.
\begin{eqnarray}
\widehat{S}^{(1)} (\bm{K}, t) &=& C^+ (\bm{K}) D^+ (\bm{K}, t)
 + C^- (\bm{K}) D^- (\bm{K}, t) \,, \\
D^{\pm} (\bm{K}, t) &=& 
\left \{
\begin{array}{lcl}
t^{-1/6} {\cal J}_{\pm \nu} (A | \bm{K} | t^{-\gamma+4/3})
& \mbox{for} & \gamma \ne \frac{4}{3} \,, \\
t^{-1/6 \pm \sqrt{25/36 - B|\bm{K}|^2}}
& \mbox{for} & \gamma  = \frac{4}{3} \,, \\
\end{array}
\right.
\end{eqnarray}
\[
A \equiv \frac{3 \sqrt{\kappa \gamma} (6 \pi G)^{(1-\gamma)/2}}{|4-3\gamma|},
~ B \equiv \frac{4}{3} \kappa (6 \pi G)^{-1/3} \,,
\]
where ${\cal J}$ means Bessel function.
$C^{\pm} (\bm{K})$ is given by the initial condition.

For the transverse mode, the solutions are same as for dust model:
\begin{equation}
{\bf s}^{T (1)} \propto \mbox{const.}, t^{-1/3} \,.
\end{equation}
The transverse mode does not have a growing solution. Therefore,
in first-order approximation, the longitudinal mode dominate
during evolution. Hereafter we consider only the longitudinal
mode solutions for the first-order solutions.

From Eqs.~(\ref{eqn:L-longi-eqn}) and (\ref{eqn:L-trans-eqn}),
we obtain the second-order perturbative equations. For
the longitudinal mode, the equation becomes
\begin{equation} \label{eqn:Lagrange-2ndL}
\nabla^2 \left ( \ddot{S}^{(2)} + 2 \frac{\dot{a}}{a}
 \dot{S}^{(2)} - 4\pi G \rho_b S^{(2)}
  - \frac{\kappa \gamma \rho_b^{\gamma-1}}{a^2} \nabla^2 S^{(2)}
 \right ) = Q^{L (2)} \,, 
\end{equation}
\begin{eqnarray}
Q^{L (2)} &=& 2 \pi G \rho_b \left [ S^{(1)}_{,ij} S^{(1)}_{,ij}
 - \left ( \nabla^2 S^{(1)} \right )^2 \right ] 
 - \frac{\kappa \gamma^2 \rho_b^{\gamma-1}}{a^2}
 \left ( \nabla^2 S^{(1)}_{,i} \nabla^2 S^{(1)}_{,i} \right )
 \nonumber \\
&& - \frac{\kappa \gamma (\gamma-1) \rho_b^{\gamma-1}}{a^2}
 \left ( \nabla^2 \nabla^2 S^{(1)} \nabla^2 S^{(1)} \right )
 - \frac{\kappa \gamma \rho_b^{\gamma-1}}{a^2}
 \left ( S^{(1)}_{,ijk} S^{(1)}_{,ijk} + 2 S^{(1)}_{,ij}
  \nabla^2 S^{(1)}_{,ij} \right ) \,.
\end{eqnarray}
For the transverse mode, after some arrangement, we can describe
as follows:
\begin{equation}
\nabla^2 \left ( \ddot{s}^{T (2)}_{,i}
 + 2 \frac{\dot{a}}{a} \dot{s}^{T (2)}_{,i} \right ) 
= Q_i^{T (2)} \,,
\end{equation}
\begin{equation}
Q_i^{T (2)} = \frac{\kappa \gamma \rho_b^{\gamma-1}}{a^2}
 \left ( S^{(1)}_{,ijk} \nabla^2 S^{(1)}_{,jk} + S^{(1)}_{,ij}
 \nabla^2 \nabla^2 S^{(1)}_{,j}
 - \nabla^2 S^{(1)}_{,j}
 \nabla^2 S^{(1)}_{,ij} - S^{(1)}_{,jk} \nabla^2 S^{(1)}_{,ijk}
 \right ) \,.
\end{equation}
Here we notice for the second-order transverse mode solutions.
In the pressure model, even if we consider
only the longitudinal mode for the first-order,
the second-order perturbation for the transverse mode
appears. In dust model, it does not appear.
Therefore, when we derive the third-order perturbative
solutions, we must consider the second-order
transverse mode.

The second-order solutions are formally written as follows:
\begin{eqnarray}
\widehat{S}^{(2)} &=& -\frac{1}{|\bm{K}|^2}
 \int^t {\rm d} t' G(\bm{K}, t, t')
 \widehat{Q}^{L (2)} (\bm{K}, t) \,, \\
\widehat{s}^{T (2)} &=& -\frac{1}{|\bm{K}|^2}
 \int^t {\rm d} t' G^T(t, t')
 \widehat{Q}^{T (2)} (\bm{K}, t) \,, \\
G^T (t, t') &=& 3 (t'- t^{-1/3} t'^{4/3} ) \,.
\end{eqnarray}
$G^L (\bm{K}, t, t')$ depends on the equation of state. If $\gamma \ne 4/3$
and $\nu = 5/(8-6\gamma)$ is not an integer, we have
\begin{eqnarray}
G^L (\bm{K}, t, t') &=& -\frac{\pi}{2 \sin \nu \pi}
 \left (-\gamma + \frac{4}{3} \right )^{-1} t^{-1/6} t'^{7/6}
 \left [ \mathcal{J}_{-\nu} \left (A |\bm{K}| t^{-\gamma+4/3} \right )
  \mathcal{J}_{\nu} \left (A |\bm{K}| t'^{-\gamma+4/3} \right )
 \right . \nonumber \\
 &&~~ \left. - \mathcal{J}_{\nu} \left (A |\bm{K}| t^{-\gamma+4/3} \right )
  \mathcal{J}_{-\nu} \left (A |\bm{K}| t'^{-\gamma+4/3} \right )
 \right ] \,,
\end{eqnarray}
and if $\gamma=4/3$,
\begin{eqnarray}
G^L (\bm{K}, t, t') &=& -\frac{1}{2 b(\bm{K})} t^{-1/6} t'^{7/6}
 \left (t^{-b(\bm{K})} t'^{b(\bm{K})} 
 - t^{b(\bm{K})} t'^{-b(\bm{K})} \right ) \,, \\
b(\bm{K}) & \equiv & \sqrt{\frac{25}{36} - B | \bm{K} |^2} \,.
\end{eqnarray}

The second-order perturbative solution for the case
of $\gamma=4/3$ in the E-dS universe model is described by
\begin{eqnarray}
\hat{S}^{(2)} (\bm{K}) &=& 
 \int {\rm d} \bm{K}' ~\Xi^L (\bm{K}, \bm{K}')
 E^L (\bm{K}, \bm{K}', t) \,, \\
\hat{s}_j^{T (2)} (\bm{K}) &=& 
 \int {\rm d} {\bf K}' ~\Xi^T_j (\bm{K}, \bm{K}')
 E^T (\bm{K}, \bm{K}', t) \,,
\end{eqnarray}
\begin{eqnarray}
\Xi^L (\bm{K}, \bm{K}') &=& -\frac{1}{(2 \pi)^3} \frac{1}{|\bm{K}|^2}
\left [
 \frac{1}{3} \left \{ (\bm{K}' \cdot (\bm{K}-\bm{K}'))^2
 - |\bm{K}'|^2 |\bm{K}-\bm{K}'|^2 \right \} \right . \nonumber \\
&& ~~ \left . + B \left \{ \frac{4}{3}
  |\bm{K}'|^2 |\bm{K}-\bm{K}'|^2
  (\bm{K}' \cdot (\bm{K}-\bm{K}'))
  + \frac{1}{3} |\bm{K}'|^4 |\bm{K}-\bm{K}'|^2 
  \right . \right . \nonumber \\
&& ~~~~~~~~~~~~ \left . \left . 
  + \left ( (\bm{K}' \cdot (\bm{K}-\bm{K}'))^3
  + 2 |\bm{K}-\bm{K}'|^2 (\bm{K}' \cdot (\bm{K}-\bm{K}'))^2
  \right ) \right \}  \right ] \,. \\
\Xi^T_j (\bm{K}, \bm{K}') &=& - \frac{i}{(2 \pi)^3} \frac{B}{|\bm{K}|^2}
 |\bm{K}-\bm{K}'|^2 (\bm{K}' \cdot (\bm{K}-\bm{K}'))
 \nonumber \\
&& ~~~~
 \left [ (\bm{K}' \cdot (\bm{K}-\bm{K}')) K'_j
  + |\bm{K}-\bm{K}'|^2 K'_j
  - |\bm{K}'|^2 (K_j - K'_j)
  - (\bm{K}' \cdot (\bm{K}-\bm{K}')) (K_j - K'_j)
 \right ] \,,
\end{eqnarray}
\begin{eqnarray}
E^L (\bm{K}, \bm{K}', t) &=& \frac{1}{2 b(\bm{K})}
 \sum_{(\oplus=\pm)} \sum_{(\otimes=\pm)} \left [
    \frac{C^{\oplus} (\bm{K}') C^{\otimes} (\bm{K}-\bm{K}')
     t^{-1/3 \oplus b(\bm{K}') \otimes b(\bm{K}-\bm{K}')}}
    {((\oplus b(\bm{K}') \otimes b(\bm{K}-\bm{K}')-\frac{1}{6})^2
     - b(\bm{K})^2)} \right ] \,, \\
E^T  (\bm{K}, \bm{K}', t) &=& 3 \sum_{(\oplus=\pm)}
     \sum_{(\otimes=\pm)} \left [
    \frac{C^{\oplus} (\bm{K}') C^{\otimes} (\bm{K}-\bm{K}')
     t^{-1/3 \oplus b(\bm{K}') \otimes b(\bm{K}-\bm{K}')}}
    {(\oplus b(\bm{K}') \otimes b(\bm{K}-\bm{K}')-\frac{1}{3})
     (\oplus b(\bm{K}') \otimes b(\bm{K}-\bm{K}'))} \right ] \,,
\end{eqnarray}
where $\sum_{(\oplus=\pm)}$ means
\begin{equation}
\sum_{(\oplus=\pm)} (\alpha^{\oplus} \oplus \beta)
 \equiv (\alpha^+ + \beta) + (\alpha^- - \beta) \,.
\end{equation}
%
\subsection{The third-order perturbative equations} \label{subsec:3rd-eq}

The third-order perturbative equation becomes very complicated.
Here we introduce scalar quantities generated by Lagrangian perturbation.
\begin{eqnarray}
\mu_1 ({\bm A}) & \equiv & \nabla \cdot {\bm A} \,, 
 \label{eqn:mu1} \\
\mu_2 ({\bm A}, {\bm B}) & \equiv & \frac{1}{2}
 \left ( (\nabla \cdot {\bm A}) (\nabla \cdot {\bm B}) 
 - A_{i, j} B_{j, i} \right ) \,, \label{eqn:mu2AB} \\
\mu_2 ({\bm A}) & \equiv & \mu_2 ({\bm A}, {\bm A}) \,, 
 \label{eqn:mu2} \\
\mu_3 ({\bm A}) & \equiv & \mbox{det} \left (A_{i, j} \right ) \,,
 \label{eqn:mu3}
\end{eqnarray}
where ${\bm A}$ and ${\bm B}$ are vector quantities.

First, we show the longitudinal mode equation. As in
the second-order perturbative equation, we separate the
terms of the third-order perturbation from the others. Then
the terms consisting of the first- and the second-order
perturbations are collected to the source term $Q^{L (3)}$.
\begin{equation}
\nabla^2 \left ( \ddot{S}^{(3)} + 2 \frac{\dot{a}}{a}
 \dot{S}^{(3)} - 4\pi G \rho_b S^{(3)}
  - \frac{\kappa \gamma \rho_b^{\gamma-1}}{a^2} \nabla^2 S^{(3)}
 \right ) = Q^{L (3)} \,. \label{eqn:3rd-L-eq}
\end{equation}
We consider the source term. Using
Eqs.~(\ref{eqn:mu1})-(\ref{eqn:mu3}), the terms are written as
follows:
\begin{eqnarray}
Q^{L (3)} &=& 4 \pi G \rho_b \left [ \mu_1 (S^{(1)}_{,i})^3
  + \mu_2 (S^{(1)}_{,i}, S^{(2)}_{,i} + s^{T (2)}_i)
 + \mu_3 (S^{(1)}_{,i}) \right ] \nonumber \\
&& + \frac{\kappa \gamma \rho_b^{\gamma-1}}{a^2}
 \left [ \mu_2 ( S^{(1)}_{,i}, S^{(2)}_{,i} + s^{T (2)}_i )
 + \mu_3 (S^{(1)}_{,i}) \right . \nonumber \\
&& ~~~~~~~~~~ \left . - S^{(1)}_{,jk} \left \{
 \partial_j \partial_k
 \left ( \mu_1 (S^{(2)}_{,i} )
  + \mu_2 (S^{(1)}_{,i}) \right ) \right \} \right . 
 \nonumber \\
&& ~~~~~~~~~~ \left . + \gamma \partial_j
 \left \{ \mu_1 (S^{(1)}_{,i}) ~\partial_j
 \left ( \mu_1 (S^{(2)}_{,i} )
  + \mu_2 (S^{(1)}_{,i}) \right ) \right \} \right .
 \nonumber \\
&& ~~~~~~~~~~ \left . + \partial_j  \left \{
 S^{(1)}_{,jk} ~\partial_k \left ( \mu_1 (S^{(2)}_{,i})
  + \mu_2 (S^{(1)}_{,i}) \right ) \right \} \right .
 \nonumber \\
&& ~~~~~~~~~~ \left . - \left \{ (S^{(2)}_{,jk} + s^{T (2)}_{j,k} )
 ~\partial_j \partial_k \mu_1 (S^{(1)}_{,i})
 - S^{(1)}_{,jk} S^{(1)}_{,kl} ~\partial_j \partial_l
 \mu_1 (S^{(1)}_{,i}) \right \} \right . \nonumber \\
&& ~~~~~~~~~~ \left . + \partial_j \left \{
 \left (-\gamma \mu_1 (S^{(2)}_{,i})
 - \gamma \mu_2 (S^{(1)}_{,i})
 + \frac{\gamma (\gamma+1)}{2} \mu_1 (S^{(1)}_{,i})^2
 \right ) \partial_j \mu_1 (S^{(1)}_{,i}) \right \}
 \right . \nonumber \\
&& ~~~~~~~~~~ \left . + \partial_j \left \{
 - (S^{(2)}_{,jk} + s^{T (2)}_{j,k}) ~\partial_k \mu_1 (S^{(1)}_{,i})
 + S^{(1)}_{,jk} S^{(1)}_{,kl} ~\partial_l \mu_1 (S^{(1)}_{,i})
 \right \} \right . \nonumber \\
&& ~~~~~~~~~~ \left . - S^{(1)}_{,jk} \partial_k
 \left \{ -\gamma \mu_1 (S^{(1)}_{,i})
 ~\partial_j \mu_1 (S^{(1)}_{,i})
 - S^{(1)}_{,jl} ~\partial _l \mu_1 (S^{(1)}_{,i})  \right \}
 \right . \nonumber \\
&& ~~~~~~~~~~ \left . + \partial_j
 \left \{ \gamma \mu_1 (S^{(1)}_{,i}) ~S^{(1)}_{,jk}
 ~\partial_k \mu_1 (S^{(1)}_{,i}) \right \} \right ] \,.
 \label{eqn:3rd-L}
\end{eqnarray}
The transverse mode also seems complicated. However, if we
neglect the first-order transverse mode, the evolution equation
is described as
\begin{equation}
-\nabla^2 \left ( \ddot{s}^{T (3)}_i + 2 \frac{\dot{a}}{a}
 \ddot{s}^{T (3)}_i \right ) =Q_i^{T (3)} \,,
\end{equation}
\begin{eqnarray}
Q_i^{T (3)} &=& \left \{ S_{,ij}^{(1)} \left ( \ddot{s}^{(2)}_{k,j}
 + 2 \frac{\dot{a}}{a} \dot{s}^{(2)}_{k,j} \right ) \right \}_{,k}
 - \left \{ S_{,jk}^{(1)} \left ( \ddot{s}^{(2)}_{i,j}
 + 2 \frac{\dot{a}}{a} \dot{s}^{(2)}_{i,j} \right ) \right \}_{,k}
 \nonumber \\
&&  + \left \{ s_{i,j}^{(2)} \left ( 4 \pi G \rho_b S^{(1)}_{,jk}
 + \frac{\kappa \gamma \rho_b^{\gamma-1}}{a^2} \nabla^2 S^{(1)}_{,jk}
 \right ) \right \}_{,k}
 - \left \{ s_{j,k}^{(2)} \left ( 4 \pi G \rho_b S^{(1)}_{,ij}
 + \frac{\kappa \gamma \rho_b^{\gamma-1}}{a^2} \nabla^2 S^{(1)}_{,ij}
  \right )  \right \}_{,k} \nonumber \\
&& + \frac{\kappa \gamma \rho_b^{\gamma-1}}{a^2}
 \left ( \nabla^2 S_{,il}^{(1)} S_{,jk}^{(1)} S_{,jkl}^{(1)}
 + \nabla^2 S_{,il}^{(1)} \nabla^2 S_{,j}^{(1)} S_{,jl}^{(1)}
 + \nabla^2 S_{,ikl}^{(1)} S_{,jk}^{(1)} S_{,jl}^{(1)} \right . \nonumber \\
&& ~~~~~~~~ \left . - S_{,ijk}^{(1)} S_{,jl}^{(1)} \nabla^2 S_{,kl}^{(1)}
 - S_{,ij}^{(1)} S_{,jkl}^{(1)} \nabla^2 S_{,kl}^{(1)}
 - S_{,ij}^{(1)} S_{,jk}^{(1)} \nabla^2 S_{,k}^{(1)}
 \right ) \,.
\end{eqnarray}
%

\subsection{The third-order perturbative solutions --
the simplest case} \label{subsec:3rd-sol}

As in the first- and the second-order solutions, the third-order
solutions are described with the Lagrangian wavenumber. Following the
method used in the second-order solutions,
the third-order solution is given by this integration:
\begin{eqnarray}
\widehat{S}^{(3)} &=& -\frac{1}{|\bm{K}|^2}
 \int_{t_0}^t {\rm d} t' G(\bm{K}, t, t')
 \widehat{Q}^{L (3)} (\bm{K}, t) \,, \\
\widehat{s}^{T (3)} &=& -\frac{1}{|\bm{K}|^2}
 \int_{t_0}^t {\rm d} t' G^T(t, t')
 \widehat{Q}^{T (3)} (\bm{K}, t) \,,
\end{eqnarray}

Here we show the third-order perturbative solutions for simplest
case, the case of $\gamma=4/3$ in the E-dS Universe model.
In this case, the contribution of the gravitational terms and
the pressure terms become identical:
\begin{equation}
4 \pi G \rho_b, \frac{\kappa \gamma \rho_b^{\gamma-1}}{a^2}
 \propto a^{-3} \,.
\end{equation}
Here we write these terms as $M/a^3 (M=\mbox{const.})$.
From the terms described by the multiplication of 
$S^{(1)}$ and $S^{(2)}$
in $Q^L$, the time evolution of a part of the third order perturbation
is contributed a
\begin{eqnarray}
&& F_1^L (\bm{K}, \bm{K}'', t) \nonumber \\
& \equiv & \int^t {\rm d} t' G(\bm{K}, t, t') ~ \frac{M}{a(t')^3}
 \left ( 
  \widehat{S}^{(1)} (\bm{K}'', t') \widehat{S}^{(2)} (\bm{K}-\bm{K}'', t')
 \right ) \nonumber \\
&=& -\frac{B M}{4 b(\bm{K}) b(\bm{K}-\bm{K}'') |\bm{K}|^2}
 ~t^{-1/2} \int {\rm d} \bm{K}'
 \Xi^L (\bm{K}-\bm{K}'', \bm{K}'-\bm{K}'') \nonumber \\
&& \sum_{(\oplus=\pm)} \sum_{(\otimes=\pm)} \sum_{(\oslash=\pm)}
 \left [ \frac{C^{\oplus} (\bm{K}'') C^{\otimes} (\bm{K}'-\bm{K}'')
  C^{\oslash} (\bm{K}-\bm{K}')}
  {((\otimes b_2 \oslash b_3 -\frac{1}{6})^2 - b_4^2)
   ((\oplus b_1  \otimes b_2 \oslash b_3 -\frac{1}{3})^2 - b_5^2)}
  t^{\oplus b_1 \otimes b_2 \oslash b_3} \right ] \,. \label{eqn:F1L}
\end{eqnarray}
\begin{equation}
b_1 \equiv b(\bm{K}''),~ b_2 \equiv b(\bm{K}'-\bm{K}''),~
b_3 \equiv b(\bm{K}-\bm{K}'),~ b_4 \equiv b(\bm{K}-\bm{K}''),~
b_5 \equiv b(\bm{K}) \,.
\end{equation}
On the other hand, from the terms described by the triplet of 
$S^{(1)}$ in $Q^L$, time evolution of a part of the third order
perturbation is contributed as:
\begin{eqnarray}
&& F_2^L (\bm{K}, \bm{K}', \bm{K}'', t) \nonumber \\
&\equiv& \int^t {\rm d} t' G(\bm{K}, t, t') ~
 \frac{M}{a(t')^3}
 \left (
  \widehat{S}^{(1)} (\bm{K}'', t') \widehat{S}^{(1)} (\bm{K}'-\bm{K}'', t')
  \widehat{S}^{(1)} (\bm{K}-\bm{K}', t')
 \right ) \nonumber \\
&=& -\frac{M}{2 b(\bm{K}) |\bm{K}|^2} ~t^{-1/2} \cdot \nonumber \\
&& \sum_{(\oplus=\pm)} \sum_{(\otimes=\pm)} \sum_{(\oslash=\pm)}
 \left [ \frac{C^{\oplus} (\bm{K}'') C^{\otimes} (\bm{K}'-\bm{K}'')
  C^{\oslash} (\bm{K}-\bm{K}')}
   {(\oplus b_1 \ominus b_2 \oslash b_3 - \frac{1}{3})^2 - b_5^2}
  t^{\oplus b_1 \ominus b_2 \oslash b_3} \right ] \,.  \label{eqn:F2L}
\end{eqnarray}
When we consider the effect of pressure, even if
${\bm s}^{T (1)} = {\bm 0}$, 
${\bm s}^{T (2)}$ appears. Therefore, the contribution
from the multiplication of $S^{(1)}$ and 
${\bm s}^{T (2)}$ also exists.
The contribution from the multiplication of $S^{(1)}$
and ${\bm s}^{T (2)}$ is given as follows:
\begin{eqnarray}
&& F_{3~j}^L (\bm{K}, \bm{K}'', t) \nonumber \\
&\equiv& \int^t {\rm d} t' G(\bm{K}, t, t') ~ \frac{M}{a(t')^3}
 \left (
  \widehat{S}^{(1)} (\bm{K}'', t') \widehat{s}_j^{T (2)}
 (\bm{K}-\bm{K}'', t')
 \right ) \nonumber \\
&=& -\frac{3M}{2 b(\bm{K}) |\bm{K}|^2} ~t^{-1/2} \int {\rm d} \bm{K}'
 ~\Xi^T_j (\bm{K}-\bm{K}'', \bm{K}'-\bm{K}'') \nonumber \\
&& \sum_{(\oplus=\pm)} \sum_{(\otimes=\pm)} \sum_{(\oslash=\pm)}
 \left [ \frac{C^{\oplus} (\bm{K}'') C^{\otimes} (\bm{K}'-\bm{K}'')
  C^{\oslash} (\bm{K}-\bm{K}')}
  {(\otimes b_2 \oslash b_3-\frac{1}{3})(\otimes b_2 \oslash b_3)
   ((\oplus b_1 \otimes b_2
   \oslash b_3 -\frac{1}{3})^2 - b_5^2)}
   t^{\oplus b_1 \otimes b_2 \oslash b_3}
  \right ] \,. \label{eqn:F3L}
\end{eqnarray}

Next, we consider the third-order transverse mode ${\bm s}^{T (3)}$.
In the transverse mode, we must make the contribution not only
from the multiplication of longitudinal modes,
but also from
the multiplication of $S^{(1)}$ and ${\bm s}^{T (2)}$:
\begin{eqnarray}
&& F_1^T (\bm{K}, \bm{K}'', t) \nonumber \\
&\equiv& \int^t {\rm d} t' G^T(t, t') ~ \frac{M}{a(t')^3}
 \left (
  \widehat{S}^{(1)} (\bm{K}'', t') \widehat{S}^{(2)} (\bm{K}-\bm{K}'', t') 
 \right ) \nonumber \\
&=& -\frac{M}{2 b(\bm{K}-\bm{K}'') |\bm{K}|^2} ~t^{-1/2} \int {\rm d} \bm{K}'
 ~\Xi^L (\bm{K}-\bm{K}'', \bm{K}'-\bm{K}'')
 \nonumber \\
&& \sum_{(\oplus=\pm)} \sum_{(\otimes=\pm)} \sum_{(\oslash=\pm)}
 \left [ \frac{C^{\oplus} (\bm{K}'') C^{\otimes} (\bm{K}-\bm{K}')
  C^{\oslash} (\bm{K}'-\bm{K}'')}
  {(\oplus b_1 \otimes b_2 \oslash b_3 - \frac{1}{6})
   (\oplus b_1 \otimes b_2 \oslash b_3 - \frac{1}{2})
   ((\otimes b_2 \oslash b_3 -\frac{1}{6})^2 - b_4^2)}
   t^{\oplus b_1 \otimes b_2 \oslash b_3} \right ] \,. \label{eqn:F1T}
\end{eqnarray}
\begin{eqnarray}
&& F_2^T (\bm{K}, \bm{K}', \bm{K}'', t) \nonumber \\
&\equiv& \int^t {\rm d} t' G^T (t, t') ~ \frac{M}{a(t')^3}
 \left (
  \widehat{S}^{(1)} (\bm{K}'', t') \widehat{S}^{(1)} (\bm{K}'-\bm{K}'', t')
  \widehat{S}^{(1)} (\bm{K}-\bm{K}', t')
 \right ) \nonumber \\
&=& -\frac{M}{|\bm{K}|^2} ~t^{-1/2} \nonumber \\
&& ~ \sum_{(\oplus=\pm)} \sum_{(\otimes=\pm)} \sum_{(\oslash=\pm)}
 \left [\frac{C^{\oplus} (\bm{K}'') C^{\otimes} (\bm{K}-\bm{K}')
  C^{\oslash} (\bm{K}'-\bm{K}'')}
  {(\oplus b_1 \otimes b_2 \oslash b_3 - \frac{1}{6})
   (\oplus b_1 \otimes b_2 \oslash b_3 - \frac{1}{2})}
  t^{\oplus b_1 \otimes b_2 \oslash b_3} \right ] \,. \label{eqn:F2T}
\end{eqnarray}
\begin{eqnarray}
&& F_3^T (\bm{K}, \bm{K}'', t) \nonumber \\
&\equiv& \int^t {\rm d} t' G^T(t, t') ~ \frac{M}{a(t')^3}
 \left (
  \widehat{S}^{(1)} (\bm{K}'', t') \widehat{s}_j^{T (2)}
 (\bm{K}-\bm{K}'', t')
 \right ) \nonumber \\
&=& -\frac{3M}{|\bm{K}|^2} ~t^{-1/2} \int {\rm d} \bm{K}' 
 ~\Xi^T_j (\bm{K}-\bm{K}'', \bm{K}'-\bm{K}'') \nonumber \\
&& \sum_{(\oplus=\pm)} \sum_{(\otimes=\pm)} \sum_{(\oslash=\pm)}
 \left [ \frac{C^{\oplus} (\bm{K}'') C^{\otimes} (\bm{K}-\bm{K}')
  C^{\oslash} (\bm{K}'-\bm{K}'')}
  {(\otimes b_2 \oslash b_3 - \frac{1}{3})
   (\otimes b_2 \oslash b_3)
   (\otimes b_1 \oplus b_2 \oslash b_3 -\frac{1}{6})
   (\otimes b_1 \oplus b_2 \oslash b_3 -\frac{1}{2})}
   t^{\oplus b_1 \otimes b_2 \oslash b_3} \right ] \,. \label{eqn:F3T}
\end{eqnarray}
The third-order perturbative solutions for the case of $\gamma=4/3$ in
the E-dS Universe model is described with
Eqs.~(\ref{eqn:F1L})-(\ref{eqn:F2T}).
Even if we restrict ourselves to the simplest case,
the third-order perturbative solutions
become very complicated. For other case, although we can derive the solutions
following similar procedures, the solutions may become difficult to analyze
structure formation.

\section{The time evolution in the planar models}
\label{sec:1D}

In the previous section, we derived the third-order perturbative
solutions for the simplest case. Although we restricted ourselves
to the simplest case, the solutions are still complicated.
For analyses of the perturbative solutions, we introduce spacial
symmetry.
If we consider a planar model, the nonlinear terms in
the gravitational term disappear. The evolution equation
for Lagrangian displacement becomes
\begin{equation}
\ddot{\vec{s}} + 2 \frac{\dot{a}}{a} \dot{\vec{s}}
 - 4 \pi G \rho_b \vec{s} = \frac{\kappa \gamma \rho_b^{\gamma-1}}
 {a^2 (1+\vec{s}_{,1})^{\gamma+1}} \vec{s}_{,11} \,,
\label{eqn:L-1D-exact}
\end{equation}
where $\vec{s}$ is Lagrangian displacement, not longitudinal mode
potential~\cite{Adler99}. From the expansion of
Eq.~(\ref{eqn:L-1D-exact}), we obtain perturbative equations.
\begin{eqnarray}
\ddot{\vec{s}}^{(1)} + 2 \frac{\dot{a}}{a} \dot{\vec{s}}^{(1)}
 - 4 \pi G \rho_b \vec{s}^{(1)}
 -\frac{\kappa \gamma \rho_b^{\gamma-1}}{a^2} \vec{s}^{(1)}_{,11}
 &=& 0 \,, \\
\ddot{\vec{s}}^{(2)} + 2 \frac{\dot{a}}{a} \dot{\vec{s}}^{(2)}
 - 4 \pi G \rho_b \vec{s}^{(2)}
 -\frac{\kappa \gamma \rho_b^{\gamma-1}}{a^2} \vec{s}^{(2)}_{,11}
 &=& -\frac{\kappa \gamma \rho_b^{\gamma-1}}{a^2}
  (\gamma+1) \vec{s}^{(1)}_{,1} \vec{s}^{(1)}_{,11} \,, \\
\ddot{\vec{s}}^{(3)} + 2 \frac{\dot{a}}{a} \dot{\vec{s}}^{(3)}
 - 4 \pi G \rho_b \vec{s}^{(3)}
 -\frac{\kappa \gamma \rho_b^{\gamma-1}}{a^2} \vec{s}^{(3)}_{,11}
 &=& \frac{\kappa \gamma \rho_b^{\gamma-1}}{a^2}
  \left [ (\gamma+1) (\vec{s}^{(1)}_{,1} \vec{s}^{(2)}_{,11} 
 +\vec{s}^{(2)}_{,1} \vec{s}^{(1)}_{,11}) \right . \nonumber \\
 && ~~~~~~~~ \left . 
 + \frac{(\gamma+1)(\gamma+2)}{2} \left \{ (\vec{s}^{(1)}_{,1})^2
 \vec{s}^{(1)}_{,11} \right \} \right ] \,.
\end{eqnarray}
For the simplest case, i.e. the Einstein-de Sitter model and $\gamma=4/3$,
using Eqs.~(\ref{eqn:F1L}) and (\ref{eqn:F2L}),
we can describe the perturbative solution for a planar model:
\begin{eqnarray}
\widehat{\vec{s}}^{(1)} (K, t) &=& C^+ (K) t^{-1/6+b(K)}
 + C^- (K) t^{-1/6-b(K)} \,, \\
\widehat{\vec{s}}^{(2)} (K, t) &=& \frac{28 i}{9}
 \int {\rm d} K' ~(K' (K-K')^2) E^L(K, K', t) \,, \\
\widehat{\vec{s}}^{(3)} (K, t) &=& \frac{784}{81}
 \int {\rm d} K' \int {\rm d} K'' \left[ (K \cdot K'' \cdot (K-K'')) 
 (K''-K') K' \right ] \nonumber \\
&& ~~~ \sum_{(\oplus=\pm)} \sum_{(\otimes=\pm)} \sum_{(\oslash=\pm)}
 \left [ \frac{C^{\oplus} (K'') C^{\otimes} (K'-K'')
  C^{\oslash} (K-K')}
  {((\otimes b_2 \oslash b_3 -\frac{1}{6})^2 - b_4^2)
   ((\oplus b_1  \otimes b_2 \oslash b_3 -\frac{1}{3})^2 - b_5^2)}
  t^{\oplus b_1 \otimes b_2 \oslash b_3} \right ] \nonumber \\
 && + \frac{140}{27} \int {\rm d} K' \int {\rm d} K''
 \left [ K' K'' (K-K'-K'')^2 \right ] \nonumber \\
&& ~~~ \sum_{(\oplus=\pm)} \sum_{(\otimes=\pm)} \sum_{(\oslash=\pm)}
 \left [ \frac{C^{\oplus} (K'') C^{\otimes} (K'-K'')
  C^{\oslash} (K-K')}
   {(\oplus b_1 \ominus b_2 \oslash b_3 - \frac{1}{3})^2 - b_5^2}
  t^{\oplus b_1 \ominus b_2 \oslash b_3} \right ] \,.
\end{eqnarray}
Here we compute power spectra ${\mathcal P} (k) \equiv \left <
\widehat{\delta}^2 (k) \right >$ of density fluctuation.
We choose the initial spectrum index as $n=+1, 0$, and $-1$. The
initial amplitude of density fluctuation is set as
$\delta_{\mbox{max}} \simeq 10^{-3}$ at $a=1$. The Jeans wavenumber
is given by hand, $K_J = 80$. In general, the Jeans wavenumber depends on
time. However, the Jeans wavenumber will still be constant
in our calculations, because we choose the polytropic index $\gamma=4/3$.

Initial conditions for two independent quantities are required
to determine $C^{\pm}$. To determine $C^{\pm}$, we set the initial
density fluctuation $\delta_0$ and the initial peculiar velocity $v_0$
by growing mode solution in ZA. The procedure was shown by
Morita and Tatekawa~\cite{Morita01}.

In ZA, the density fluctuation diverges at
$a \simeq 1000$. Here we observe the difference between first-,
second-, and third-order approximations. Therefore, we calculate
the evolution just before shell-crossing, i.e., density divergence.

\begin{figure}
 \includegraphics{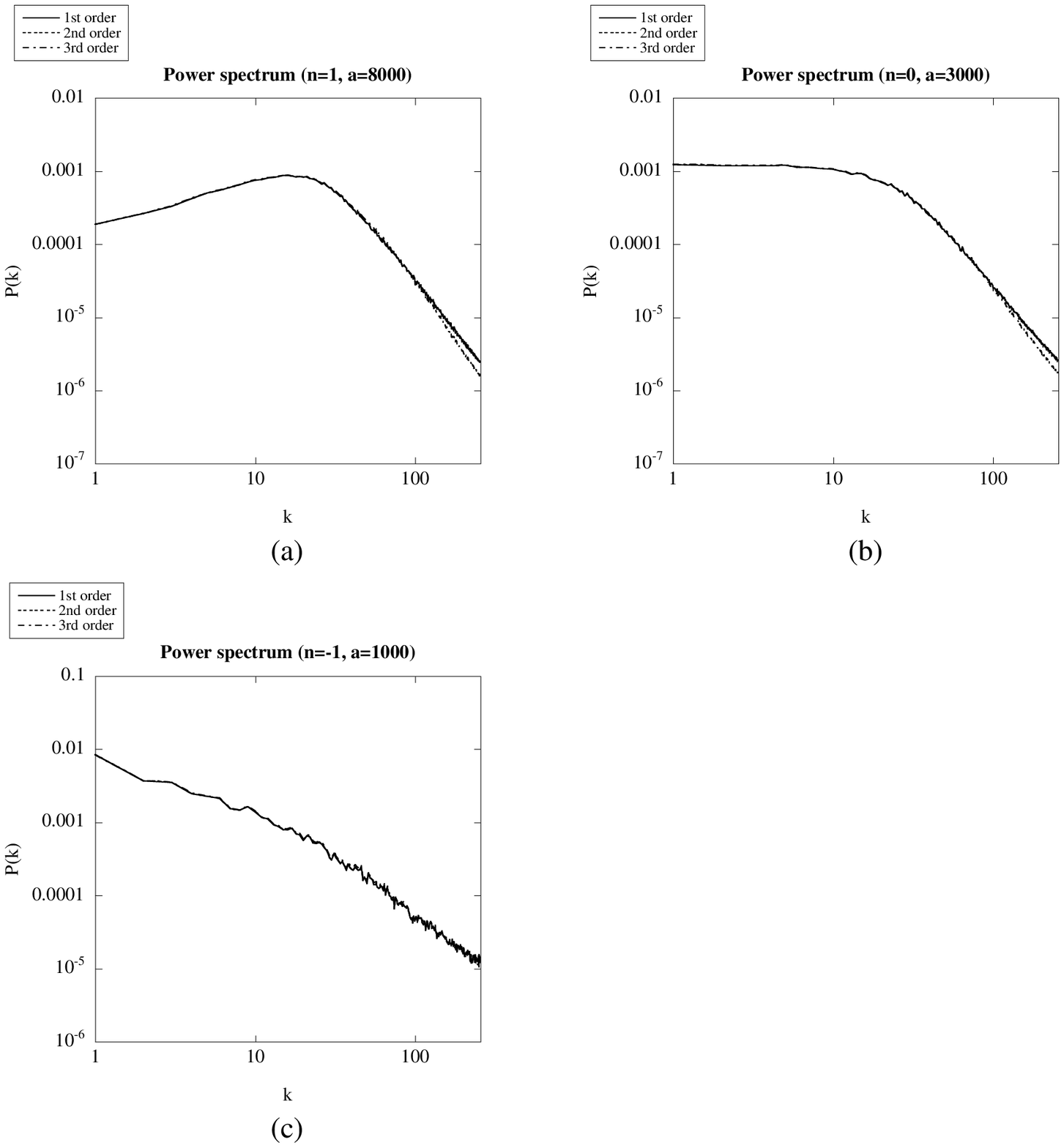}
 \caption{Power spectra of density fluctuation computed from
 Lagrangian first-, second-, and third-order approximations.
 (a) The initial spectrum index is $n=1$. The power spectra are at
 $a=8000$. The difference between the results by the first-,
 second-, and third-order Lagrangian approximation appears.
 (b) The power spectra are at $a=3000$ for $n=0$. In this case,
 the effect of the third-order approximation also appears.
 (c) The power spectra are at $a=1000$ for $n=-1$.
 The difference from Lagrangian approximation will still be very small
 just before shell-crossing.}
\label{fig:1D-spectrum}
\end{figure}

Fig.~\ref{fig:1D-spectrum} shows the power spectra for $n=+1, 0, -1$
at $a=8000 ~(n=1)$, $a=3000 ~(n=0)$, and $a=1000 ~(n=-1)$.
In these figures, we take an ensemble average of 900 samples.
For $n=1$ and $n=0$, a difference between the results by the first-,
second-, and third-order Lagrangian approximations appears.
The effect of the higher-order approximation suppresses
the evolution of the density fluctuation.
However, for $n=-1$, because the initial
power in the large wavenumber component is small, the pressure
does not surpress the evolution well. Therefore, the difference
between the Lagrangian approximations still be very small
even just before shell-crossing.

In the one-dimensional model, the pressure only affects the nonlinear effect.
However, in the three-dimensional realistic model, the gravitational force
also affects the nonlinear effect, and the difference between
first-, second-,
and third-order approximation obviously appears. In fact, according to
the comparison between the first-order approximation and full-order
numerical calculation, the difference becomes large in the strongly
nonlinear region~\cite{Tatekawa04B}.

\section{Summary and Concluding Remarks}
\label{sec:summary}

In this paper, we showed the third-order Lagrangian perturbative equations
for the cosmological fluid with pressure. Then we derived third-order
perturbative solutions for a simple case.

In our analysis for a planar model, the effect of the third-order
perturbation seems very small even at the nonlinear stage.
However, our result does not show that we can ignore the third-order
perturbation easily. As mentioned in Sec.~\ref{sec:1D}, when we
introduce planar symmetry, the nonlinear term of the gravitational
force disappears. When we consider the effect of nonlinear pressure
and gravitational force, the third-order perturbation is expected
as a powerful tool to treat high-density regions.

Recently several dark matter models have been proposed~\cite{DarkMatter}.
If the interaction in some kind of
dark matter can be described by the effective pressure, we can examine
the behavior of the density fluctuation in a quasi-nonlinear stage.
Furthermore, when we compare the observations and the structure
that is formed by using the pressure model, we can delimit
the nature of the dark matter.
Especially when we consider high-density regions, the
third-order perturbative solutions may become useful.

\begin{acknowledgments}
We are grateful to Kei-ichi Maeda for his continuous encouragement.
We would like to thank Hajime Sotani for
useful discussion and comments regarding this work.
The numerical calculation was in part carried out on the general common-use computer system at the Astronomical Data Analysis Center, ADAC, of the National Astronomical Observatory of Japan and Yukawa Institute Computer
Facility.
This work was supported by a Grant-in-Aid for Scientific
Research Fund of the Ministry of Education, Culture, Sports, Science,
and Technology (Young Scientists (B) 16740152).
\end{acknowledgments}

\appendix
\section{Second-order equation for spherical symmetric model
with pressure} \label{sec:2nd-sphe}

When we consider spherical collapse or expansion of a spherical void,
we introduce spherical symmetry. 

The second order perturbative equation (Eq.~(\ref{eqn:Lagrange-2ndL}))
is changed as follows:
\begin{eqnarray} \label{eqn:Lagrange-sphe2nd}
&& \nabla^2 \left (\ddot{\zeta} + 2 \frac{\dot{a}}{a} \dot{\zeta}
 - 4 \pi G \rho_b \zeta - \frac{1}{a^2} \frac{{\rm d} P}{{\rm d} \rho}
  (\rho_b) \nabla^2 \zeta \right ) \nonumber \\
&=& 2 \pi G \rho_b \left[ \frac{-2 S' (S' + 2r S'')}{r^2} \right ] \nonumber \\
&& -\frac{{\rm d} P}{{\rm d} \rho} (\rho_b) \left [
 \frac{(-2 S' + 2r S'' + r^2 S''')^2}{r^4} \right. \nonumber \\
&& ~~~~ \left .
  + \frac{6 (S')^2 - 12r S' S'' + 6 r^2 (S'')^2 + r^4 (S''')^2}{r^4}
   \right. \nonumber \\
&& ~~~~ \left . + 2 \frac{ S'( -4 S' + 8r S'' + 2r^2 S''')
    + r^2 S'' (-4 S''+ 2r S'''+ r^2 S'''')}{r^4}
  \right ] \nonumber \\
&& -\frac{{\rm d}^2 P}{{\rm d} \rho^2} (\rho_b) \rho_b \left [
  \frac{(4 S''' + r S'''')(2S' + r S'')}{r^2} 
  +  \frac{(-2 S' + 2r S'' + r^2 S''')^2}{r^4} \right ] \nonumber \\
&=& -4 \pi G \rho_b \frac{S' (S' + 2r S'')}{r^2} \nonumber \\
&& - \frac{1}{r^4}\frac{{\rm d} P}{{\rm d} \rho} (\rho_b) \left [
 2 (S')^2 - 4r S' S'' + 2r^2 (S'')^2 + 8r^3 S'' S'''
  + 2 r^4 ((S''')^2+ S'' S'''')
 \right ] \nonumber \\
&& -\frac{1}{r^4} \frac{{\rm d}^2 P}{{\rm d} \rho^2} (\rho_b) \rho_b
 \left [
 4 (S')^2 - 8r S' S'' + 4r^2 ((S'')^2 + S' S''') 
 + 2r^3 (S' S'''' + 4 S'' S''') + r^4 (S'' S'''' + (S''')^2)
 \right ] \,,
\end{eqnarray}
where $S$ and $\zeta$ are the first- and the second-order
Lagrangian perturbation, respectively. Because the mode-coupling
of the first-order perturbation becomes very complicated,
it seems very difficult to solve the equation~(\ref{eqn:Lagrange-sphe2nd}).

\end{document}